\documentclass[aps,pre,twocolumn,showpacs,showkeys,superscriptaddress]{revtex4}
\usepackage{graphicx}
\usepackage{dcolumn}
\usepackage{bm}

\begin{document}


\title{Spatial Growth of Real-world Networks}


\author{Marcus Kaiser}
\email[Corresponding author. Electronic address: ]{m.kaiser@iu-bremen.de}
\affiliation{International University Bremen, School of Engineering and Science, Campus Ring 6, 28759 Bremen, Germany}

\author{Claus C. Hilgetag}
\affiliation{International University Bremen, School of Engineering and Science, Campus Ring 6, 28759 Bremen, Germany}
\affiliation{Boston University, Department of Health Sciences, Boston, USA}

\date{\today}

\begin{abstract}
Many real-world networks have properties of small-world networks, with clustered local neighborhoods and low average-shortest path (ASP). They may also show a scale-free degree distribution, which can be generated by growth and preferential attachment to highly connected nodes, or hubs. However, many real-world networks consist of multiple, inter-connected clusters not normally seen in systems grown by preferential attachment, and there also exist real-world networks with a scale-free degree distribution that do not contain highly connected hubs. We describe spatial growth mechanisms, not using preferential attachment, that address both aspects.
\end{abstract}

\pacs{89.75.Hc, 89.75.Da, 89.40.Bb, 82.30.Nr}
\keywords{scale-free, small-world, spatial networks, neural networks, metabolic networks}

\maketitle


\section{Introduction}
Many real-world networks show {\it small-world} properties \cite{Watts1998}. Their average clustering coefficient, representing the proportion of direct links between the neighbors of a node, is higher than in same-size random networks, while they maintain a comparable average shortest path (ASP). 
The giant component of some of these networks has been shown to consist of several clusters, which contain strongly interlinked nodes and form only sporadic connections to other clusters. For instance, the cortical systems networks in macaque monkey and cat brains possess such a multi-cluster organization \cite{Hilgetag2000b}. Moreover, various complex linked systems have been described as {\it scale-free} networks \cite{Barabasi1999,Huberman1999}, in which the probability for a node possessing $k$ edges is  $P(k)\propto k^{-\gamma}$. It has been suggested that this large class of networks may be generated by mechanisms of growth and preferential attachment, that is, the preferred linking of new nodes to already highly connected network nodes \cite{Barabasi1999}. An essential aspect of many real-world networks is, however, that they exist and develop in metric space. Therefore, questions arise how nodes are able to identify highly connected distant hubs and why they would attach to them, rather than to nearby nodes \cite{Caldarelli2002}. Moreover, long-range connections to hubs violate optimal wiring principles \cite{Cherniak1994}. For example, a city in New England would normally consider constructing a new highway to nearby Boston, rather than to faraway Los Angeles, even if Los Angeles represents a larger hub in the US highway system. 

Previous spatial growth algorithms, in which the probability for edge formation decreased with node distance, predetermined the position of all nodes at the outset \cite{Waxman1988,Yook2002}. Starting with the complete set of nodes, which was distributed randomly on a spatial grid, connections were established depending on distance \cite{Watts1999,Kleinberg2000,Eames2002}. Additionally, connected nodes could be drawn together by a {\it posteriori} pulling algorithm, which resulted in spatial clusters of connected nodes \cite{Segev2003}.  Such mechanisms, however, appear unsuited as a general explanation for growing biological and artificial systems with newly forming nodes and connections. 

\section{Spatial Network Development Algorithm}
In an alternative approach, we employed a model of spatial growth in which the nodes, their positions and connections were established during development. Starting with one node at the central position (0.5; 0.5) of the square embedding space (edge length one), the following algorithm was used: \\
1) A new node position was chosen randomly in two-dimensional space with coordinates in the interval [0; 1]. \\
2) Connections of the new node, $u$, with each existing node, $v$, were established with probability 
\begin{equation}\label{exponential}
P(u,v) = \beta \ e^{-\alpha \ d(u, v)},
\end{equation}
where $d(u, v)$ was the spatial (Euclidian) distance between the node positions, and $\alpha$ and $\beta$ were scaling coefficients shaping the connection probability \cite{Waxman1988}. \\
3) If the new node did not manage to establish connections, it was removed from the network. In that way, newly forming nodes could only be integrated within the vicinity of the existing network, making the survival of new nodes dependent on the spatial layout of the present nodes. \\
4) The algorithm continued with the first step, until a desired number of nodes was reached.\\
Parameter $\beta$ ("density") served to adjust the general probability of edge formation and was chosen from the interval [0; 1]. The nonnegative coefficient $\alpha$ ("spatial range") exponentially regulated the dependence of edge formation on the distance to existing nodes. Such spatial constraints are present during the development of many real networks. In biological systems, for instance, gradients of chemical concentrations, or molecule interactions decay exponentially with distance \cite{Murray1990}. 

The algorithm allowed some nodes to be established distant to the existing network, although with low probability. Subsequent nodes placed near to such 'pioneer' nodes would establish connections to them and thereby generate new highly-connected regions away from the rest of the network. Through this mechanism multiple clusters were able to arise, resulting in networks in which nodes were clustered topologically as well as spatially.

In a slightly modified approach the growth model could employ a power-law to describe the dependence of edge formation on the spatial distance of nodes: 
\begin{equation}\label{powerlaw}
P(u,v) = \sigma \ d(u,v)^{-\tau}.
\end{equation}
By this mechanism the probability of establishing distant nodes would be increased even further. For example, simulating networks of similar size (50 networks; $n=100$; $density=0.04$; square embedding space edge length 100) for both types of distance dependencies, the power-law (Eq.~\ref{powerlaw}, $\sigma=1$, $\tau=1$) resulted in higher total wiring length (6303) compared to networks generated by exponential edge probability (Eq.~\ref{exponential}, $\alpha=0.35$, $\beta=1$, total wiring length 1077 units). In the following investigations, however, we concentrated on the exponential approach outlined above, since our simulations indicated that power-law edge probability was unable to yield small-world networks (tested parameter ranges $\sigma \in [0.004; 2]$ and $\tau \in [0.125; 64]$).

Another essential network feature investigated in the model was the presence or absence of hard spatial borders that limit network growth. Borders occur in many compartmentalized systems, be it mountains or water surrounding geographical regions, cellular membranes separating biochemical reaction spaces, or the skull limiting expansion of the brain. Depending on coefficient $\alpha$ and the network size, our simulated networks never reached a hard border ('virtually unlimited growth'), or quickly arrived at the spatial limits, so that new nodes could then only be established inside the existing networks. Naturally, virtually unlimited growth would eventually also arrive at the hard borders, after sufficiently sustained network growth. However, in the context of our simulations, growth could be considered virtually unlimited if for a chosen network size at the end of the algorithm all nodes were still far away from the borders (by at least 0.25 units).

In the following, we describe different types of spatially grown networks resulting from low or high settings for parameters $\alpha$ and $\beta$, and present examples of real-world networks corresponding to the generated types.

For the generated networks, two network properties are shown, which have been used previously to characterize complex networks \cite{Watts1998}. The average shortest path (ASP, similar, though not identical, to characteristic path length $L$ \cite{Watts1999}) of a network with $N$ nodes is the average number of edges that has to be crossed on the shortest path from any one node to another.
\begin{equation}\label{asp}
ASP = {1 \over N (N-1)} \sum_{i, j} d(i,j)   \ \ \ \ \ with \ i\ne j,
\end{equation}
where $d(i,j)$ is the length of the shortest path between nodes $i$ and $j$.
The clustering coefficient of one node $v$ with $k_v$ neighbors is 
\begin{equation}\label{clustercoef}
C_v = {|E(\Gamma_v)| \over {k_v \choose 2} },
\end{equation}
where $|E(\Gamma_v)|$ is the number of edges in the neighborhood of $v$ and ${k_v \choose 2}$ is the number of possible edges \cite{Watts1999}. In the following analyses we use the term clustering coefficient as the average clustering coefficient for all nodes of a network.

Algorithms for network generation, calculation of network parameters and visualization were developed in Matlab (Release 12, MathWorks Inc., Natick) and also implemented in C for larger networks. For each parameter set and network size, 50 simulated networks were generated and analyzed (20 in the case of virtually unlimited growth, due to computational constraints).

\section{Modeled Types of Networks}

\subsection{Sparse Networks (limited and virtually unlimited growth).} 
\begin{figure}
 \includegraphics{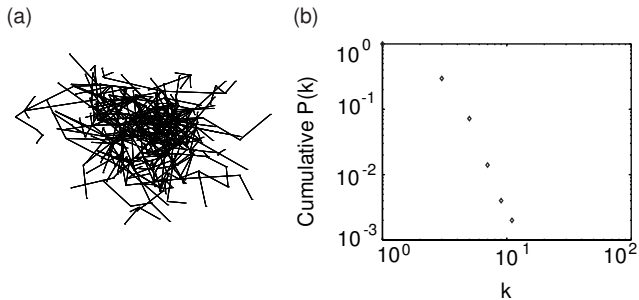}
 \caption{\label{sf} (a) Sparse network (density 0.42\%) with 500 nodes obtained by limited growth ($\alpha=2$, $\beta=0.001$). (b) Cumulative degree probability $P(k)$ that a node possesses $k$ edges for the network shown in (a). A power-law of the degree distribution ($\gamma=2.43$) can be observed.}
 \end{figure}

For very small $\beta$ ($<0.01$), sparse networks were generated (Fig. \ref{sf}a) in which only a small proportion of all possible edges was established. The resulting networks were highly linear, that is, exhibiting one-dimensional chains of nodes, independent of limited or virtually unlimited growth (parameter $\alpha$). The histograms of chain-lengths found in these networks, indicating the number of nodes in the chains, were similar to those of random networks with the same density. Unlike in random networks, however, the clustering coefficient was lower than the network density, and despite lacking clusters and hubs with large degree $k$, these networks possessed a power-law degree distribution, with high ASP (Fig. \ref{sf}b). The power-law exponent was small, in the range of [1.7; 2.1]; and in the simulated networks of 100 nodes the cut-off for the maximum degree of the scale-free networks was 16. Given their low maximum degree, these networks with low clustering and long linear chains of nodes could be called linear scale-free. 

{\bf Example: German Highway System.} We identified a linear scale-free organization in the German highway ("Autobahn") system. The highway network of 1,168 nodes was compiled from data of the 'Autobahn-Informations-System' \cite{auto}. The ratio of clustering coefficient and density of the highway system, which can be seen as a linearity coefficient, was 0.64. This system is also an example for a scale-free ($\gamma=2.8$), yet not small-world, network, as its ASP was twice as large as for comparable random networks. 

A similar type of organization was also found for scale-free protein-protein interaction networks \cite{Jeong2001} ($k_{max}\approx 20$).

\subsection{Dense Networks (limited and virtually unlimited growth).} 
For higher edge probability ($\beta \rightarrow 1$), a noteworthy difference between limited and virtually unlimited growth became apparent. While it was impossible to generate high network density under virtually unlimited growth conditions, the introduction of spatial limits resulted in high density and clustering, as well as low ASP. This was due to the fact that, in the virtually unlimited case, new nodes at the borders of the existing network were surrounded by fewer nodes and therefore formed fewer edges than central nodes within the network. In the limited case, however, the network occupied the whole area of accessible positions. Therefore, new nodes could only be established within a region already dense with nodes and would form many connections. 

\begin{figure}
 \includegraphics{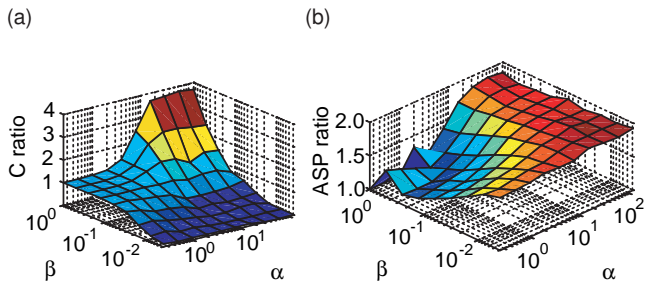}
 \caption{\label{sw} Comparison of small-world properties of spatial and random networks for N=100 nodes. Each data point represents the average for 50 networks. (a) Ratio of the clustering coefficient C of the generated networks divided by the clustering coefficient for comparable random networks. A large ratio is one feature of small-world networks. (b) Ratio of the average-shortest paths, ASP, of spatial-growth and comparable random networks.}
 \end{figure}

Figure \ref{sw} shows the relation between small-world graph properties and growth parameters $\alpha$ and $\beta$ for networks consisting of 100 nodes. The ratio of the clustering coefficient in spatial growth compared to random networks was larger than one (indicating small world graphs), if the values for $\alpha$ and $\beta$ were high (Fig. \ref{sw}a). The ASP in the generated networks normalized by the ASP in random networks with similar density was similar for low values of $\alpha$ and high values of $\beta$. For these networks the likelihood of edge formation was high and --- because of the low value of $\alpha$ --- independent from spatial distance. Such networks resembled random growth, with the clustering coefficient possessing the same value as the density ($C/C_{random}\approx 1$).

\begin{figure}
 \includegraphics{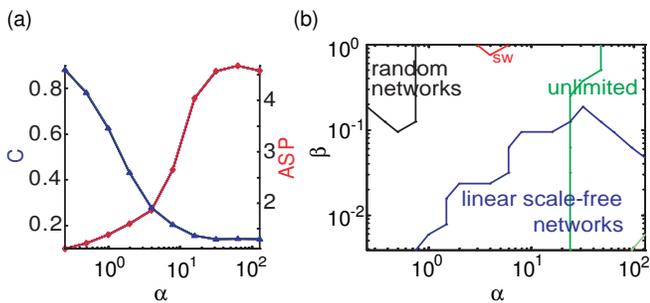}
 \caption{\label{overview}Exploration of model parameter space. (a) For dense networks ($\beta=1$, $N=100$ nodes), an increased dependence of edge formation on distance (parameter $\alpha$) led to an increase of $ASP$ (diamonds) and a decrease in clustering coefficient $C$ (triangles). (b) Overview of network types for different spatial growth parameters ($N=100$ nodes). Low values of $\alpha$ made edge formation independent from distance and resulted in random networks (black). For large values of $\alpha$ only nodes near the existing network could establish connections, and the hard borders were not reached (virtually unlimited, green). The area labeled linear scale-free (blue) was a region in which sparse and highly linear networks showing a scale-free degree distribution occurred. Only a small part of the parameter space (red) showed properties of small-world networks. }
 \end{figure}

In a small interval of intermediate values for $\alpha$ ($\alpha\approx 4$, $\beta=1$), networks 
exhibited properties of small-world networks (ASP and clustering coefficient shown in Fig. \ref{overview}a). Here, the ASP was comparable to that in random networks of the same size ($ASP\approx ASP_{random}$), while the clustering coefficient was 39\% higher than in random networks \cite[p. 114]{Watts1999}. An overview of the parameter space and the resulting random, small-world, virtually unlimited or linear scale-free networks is given in Figure \ref{overview}b.

{\bf Example: Cortical Connectivity.} One biological example for small-world spatial networks with high clustering coefficient and high density are the well studied, clustered systems of long-range cortical connectivity in the cat and macaque monkey brains \cite{Scannell1999,Young1993,Hilgetag2000b}. We employed the model in order to generate networks with identical number of nodes and edges and comparable small-world properties. 
While small-world networks could be generated in the appropriate parameter range of the model (Fig. \ref{overview}b), the biological networks featured even stronger clustering. We found, however, that such networks could be produced by extending the local range of high connection probability, so that $P=1$ for $distance_{cat}<0.18$, $distance_{macaque}<0.11$ and $P$ decaying exponentially as before for larger distances (this was implemented by setting $\alpha_{cat}=5$, $\alpha_{macaque}=8$ and for both networks $\beta=2.5$ and thresholding probabilities larger than one to one). The modified approach therefore combined specific features of the biological networks with the general model of limited spatial growth. This yielded networks with distributed, multiple clusters, and average densities of around 30\%  (for simulated cat brain connectivity) and 16\% (monkey connectivity).  Moreover, these networks had clustering coefficients of 50\% and 40\%, respectively, very similar to the biological brain networks \cite{Hilgetag2000b}, as shown in Table \ref{brain}.  

Comparison of the biological and simulated degree distributions, moreover, showed a significant correlation (Spearman's rank correlation $\rho$ = 0.77 for the cat network, $P<3\times10^{-3}$; and $\rho$ = 0.9 for the macaque network, $P<2\times10^{-5}$). On the other hand, the BA-model \cite{Barabasi1999}, using growth and preferential attachment, yielded similar densities and clustering coefficients, but was unable to generate multiple clusters as found in the real cortical networks.

\begin{table}
\caption{Comparison of cortical and simulated networks. Shown are the clustering coefficient $C_{cortical}$ of cortical networks of cat and macaque with a given number of nodes $n$ and density $d$ as well as the clustering coefficient $C_{spatial \ growth}$ of generated networks with identical node number and similar density.\label{brain}}
\begin{ruledtabular}
\begin{tabular}{lllll}
\ \		& $n$	& $d$	& $C_{cortical}$ &	$C_{spatial \ growth}$ \\
\hline \\
cat		& 55	& 0.30	& 0.55		  & 	0.5 \\
macaque		& 73	& 0.16	& 0.46		  & 	0.4 \\
\end{tabular}
\end{ruledtabular}
\end{table}

In contrast to limited growth, virtually unlimited growth simulations with high $\beta$ resulted in inhomogeneous networks with dense cores and sparser periphery. It is difficult to imagine realistic examples for strictly unlimited development, as all spatial networks eventually face internal or external constraints that confine growth, may it be geographical borders or limits of their energetic and material resources. However, virtually unlimited growth may be a good approximation for the early development of networks before reaching borders.

\section{Classifying Types of Network Development}
\begin{figure}
 \includegraphics{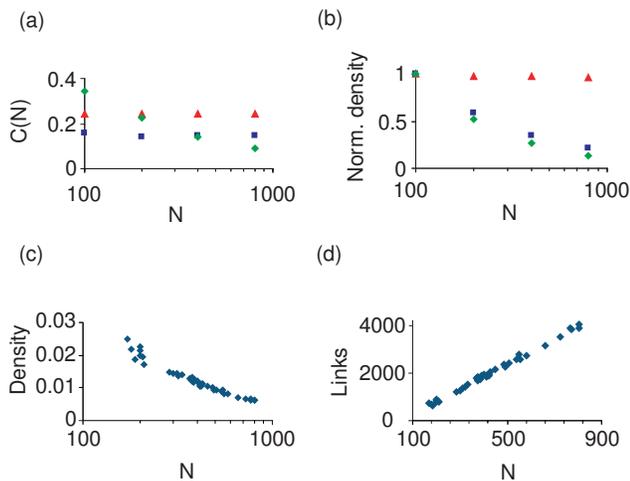}
 \caption{\label{dev}Comparison of the dependence of clustering coefficient $C(N)$ and density on network size (number of nodes, $N$). (a) For the simulated networks the clustering coefficient remained constant for limited (triangles, $\alpha=5$, $\beta=1$) and virtually unlimited (boxes, $\alpha=200$, $\beta=1$) spatial growth, but decreased for growth with preferential attachment (diamonds). (b) Density was independent of network size only for limited spatial growth. (c) Density depending on network size ($N$) for the metabolic networks of 43 different organisms (15). (d) A critical measure for network development was the dependence of network size on the number of links. For metabolic networks, this relationship was strongly linear. }
 \end{figure}

Different network growth types can be distinguished by assessing the evolution of network density and clustering coefficient. Growth with preferential attachment as well as spatial growth lead to clustering coefficients, $C(N)$, that depend on the current size of the network, that is, the number of nodes, $N$ (Fig. \ref{dev}a). While $C(N)$ decreases with network size for networks generated by the BA-Model \cite{Barabasi1999}, it remains constant for spatial-growth networks. Virtually unlimited or limited spatial growth can thus be distinguished, since density decreases with network size for unlimited growth, while remaining constant for limited growth (Fig. \ref{dev}b). 

{\bf Example: Evolution of metabolic networks.} We applied this concept to classifying the development of real-world biological networks. The evolution of metabolic systems, for instance, can be seen as an incorporation of new substances and their metabolic interactions into an existing reaction network. Reviewing 43 metabolic networks in species of different organizational level \cite{Barabasi2000b}, the clustering coefficient of these systems remained constant across the scale \cite{Ravasz2002}, whereas their density (Fig. \ref{dev}c) decreased with network size. This indicated features of virtually unlimited network growth. The relation between the number of links and nodes in these systems was linear (Fig. \ref{dev}d), with a slope of 5.2, so that the number of interactions of a metabolite was not increasing with network size. Such linear growth may ensure that the metabolic systems remain connected (with the number of reactions larger than substances, as a necessary condition for connectedness), while not becoming too complex too quickly (as, for instance, with exponential addition of new reactions). 

\section{Conclusions}
We have proposed a new kind of spatial growth mechanism, incorporating both limited and virtually unlimited growth, that can produce a variety of metric real-world networks. The metric is not limited to Euclidian space as in the discussed examples, but may also use measures of similarity to define the link probability (e.g., social relations, \cite{Watts2002}).

In contrast to previously studied spatial graphs \cite{Watts1999}, networks generated by our model were always connected. Moreover, the approach was able to generate small-world graphs, which is thought not to be possible in the spatial graph model in which positions are chosen randomly {\it before} edge formation \cite{Watts1999}. Finally, the model was also able to produce scale-free networks with relatively low maximum degree, similar to, for example, the German highway system. 

A systematic evaluation of model parameter space was carried out at the specific network size of 100 nodes, which was feasible computationally. It would be interesting to also evaluate larger or smaller network sizes and to investigate for them, if small-world networks can be generated in a larger range of parameters $\alpha$ and $\beta$.

Several algorithms have been proposed for the generation of different types of topological networks, in which links do not reflect physical distances, but merely the connectivity of the system \cite{Watts1998,Barabasi1999,Newman2001}. Examples for such networks include the World-Wide Web, financial transaction networks, and, to some extent, networks of airline transportation. The present model extends previous approaches to the development of spatial networks, such as cellular and brain connectivity networks, or food webs and many systems of social interactions. Spatial as well as temporal constraints shape network growth, and intrinsic or external spatial limits may determine essential features of the structural organization of linked systems, such as clustering and scaling properties. Borders, for instance, appear to have been critical for early chemical evolution, ensuring clustering of good replicators and preventing the spreading of short templates with limited replication function \cite{Szabo2002}. The same applies to cortical networks where elimination of growth limits results in a distorted network topology \cite{Kuida1998}.

The specific spatio-temporal conditions for the development of different types of real-world networks warrant further investigation. They may be of additional interest, as local spatial growth mechanisms also imply global optimization of path lengths in connected systems \cite{Valverde2002}. 	\\

\begin{acknowledgments}
We thank N. Sachs, H. Jaeger, M. Zacharias and A. Birk for critical comments on the manuscript.
\end{acknowledgments}


\newcommand{\noopsort}[1]{} \newcommand{\printfirst}[2]{#1}
  \newcommand{\singleletter}[1]{#1} \newcommand{\switchargs}[2]{#2#1}

\end{document}